# Phase diagram of hard board like colloids from computer simulations

*Stavros D. Peroukidis*[*,†] *and Alexandros G. Vanakaras*[†]

Department of Materials Science, University of Patras, Patras 26504, Greece

ABSTRACT

The rich mesophase polymorhism and the phase sequence of board-like colloids depends critically on their shape anisometry. Implementing extensive Monte Carlo simulations, we calculated the full phase diagram of sterically interacting board-like particles, for a range of experimentally accessible molecular dimensions/anisometries of colloids of this shape. A variety of self organized mesophases including uniaxial and biaxial nematics, smectic, cubatic and columnar phases have been identified. Our results demonstrate clearly that the molecular anisometry influences critically not only the structure and the symmetry of the mesophases but also, and perhaps more interestingly, the phase sequence among them. New classes of phase sequences such as nematic-nematic and, for the first time, a direct transition from a discotic and a biaxial nematic to an orthogonal smectic A phase have been identified. The molecular geometry requirements for such a phase behavior have been located.

---

[*] To whom the correspondence should be adressed. E-mail: peroukid@upatras.gr

[†] Department of Materials Science, University of Patras, Patras 26504, Greece

Colloidal suspensions of sterically interacting anisometric particles exhibit a rich variety of mesophases and phase transformations among them.[1-6] The local structure and eventually the macroscopic behavior of these self-organized anisotropic fluids are entropy driven since they are determined mainly by excluded volume interactions.[1,7-9] Recently, colloidal dispersions of board like goethite particles were demonstrated[5,10] to exhibit, in addition to the usual uniaxial nematic phases, a spontaneously formed stable biaxial nematic phase as well as a series of nematic-nematic ($N-N$) phase transitions under the influence of external fields.[11] At high concentrations these systems exhibit smectic and/or columnar phases with structures not yet unambiguously indentified.

The possibility of formation of stable biaxial nematic phases by $D_{2h}$-symmetric hard board-like particles having the so called "dual shape" i.e. particles having almost equal length to width and width to thickness ratios, was predicted theoretically four decades ago by Straley.[12] Surprisingly, since then there are no computer simulation studies of this particularly simple model with the exception of the simulations by Bates et. al.[13] In these simulations, however, the orientation of the long axis of the particles was fully oriented, thus preventing the possibility of an orientationally isotropic fluid. On the other hand it has been shown that hard biaxial ellipsoids of certain anisometries exhibit stable uniaxial and biaxial nematic phases.[14] However, as it has been pointed out by Bolhuis et. al.[15], this system is unlikely to exhibit positionaly ordered liquid crystalline (LC) phases and indeed it does not[14]. This is an inherent weakness of the hard ellipsoid model, preventing the study of the relative stability of the nematic state with respect to the positionaly ordered LC states, which are always present in the phase sequence of real board-like colloidal suspensions.[5,10]

To overcome the weaknesses of the aforementioned models, we model the board like particles as hard spheroplatelets (SPs), a tractable model introduced by Mulder[16] for analytical calculations as a generalization of the well studied spherocylinder model. A SP particle consists of a rectangular box with dimensions $(l-d)\times(w-d)\times d$, capped at its corners by quarter spheres of diameter $d$ and half cylinders with diameter $d$ and length $(l-d)$ and $(w-d)$ forming a convex body of orthorhombic symmetry (see Figure 1). Without loss of generality we assume that $l \geq w \geq d$ and assign a body fixed frame as shown in Figure 1. The dimensionless length and width are given by $l^* = l/d$ and $w^* = w/d$ respectively. When $w^* = 1$ the SP particle becomes a spherocylinder with aspect ratio $l^*-1$; a system whose phase diagram is well known, for a wide range of aspect ratios, from the seminal works of Bolhuis et. al. [15] and McGrother et. al.[17] In the other $w^* = l^*$ limit the SP particle transforms into a tetragonal spheroplatelet ($D_{4h}$-symmetry) and becomes very similar with the tetragonal parallelepiped particles simulated by John et. al.[18]

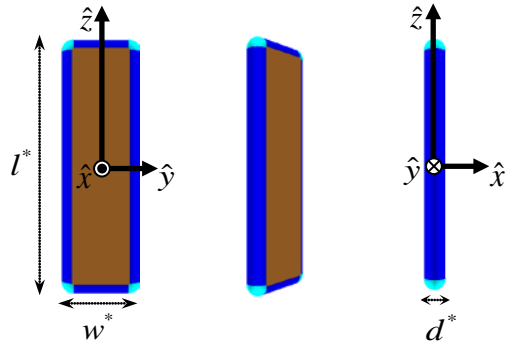

**Figure 1.** Views of the simulated hard board spheroplatelet particles.

The phase behavior and the molecular organization of the SPs is studied here by Metropolis Monte Carlo computer simulations in the isothermal isobaric ensemble ($NpT$) using variable size simulation boxes with periodic boundary conditions.[19] Most of the simulations were performed for $N \approx 1000$ particles although much larger systems were simulated in certain cases. The pressure ($p^*$) versus packing fraction ($\eta$) equation of state (EOS) of each simulated system in the ($l^*, w^*$) space, was calculated with compression runs from the isotropic phase. Here $\eta = N\upsilon/\bar{V}$ is the packing fraction, $\upsilon$ is the molecular volume, $p^* = p\upsilon/k_B T$ is the reduced pressure and $\bar{V}$ is the ensemble average of the volume of the simulated system at $p^*$. Expansion series from highly packed ordered phases were performed as well to check whether the observed phase sequence on compression includes metastable (over-compressed) states. Since our prime interest in this letter is on the LC behavior of the SP systems, we present results for particles with aspect ratios $w^*$ and $l^*$ in the range $5 \leq l^* < 12$ and $1 < w^* < 12$. Particles with aspect ratio $l^* < 5$ freeze directly into crystalline phases or into long lived glassy states. Systems with $l^* = 5$ and $w^* < 2$ exhibit, in accordance with previous studies, a rather narrow range of stable smectic-A (*SmA*) ordering before they crystallize;[15,17] mesomorphism is removed from the phase sequence when $w^* > 2$.

To trace the phase transition boundaries between the isotropic and the nematic phase we have identified the orientational symmetry of the phases and the associated order parameters through diagonalization of the order tensors,[20] $\mathbf{Q}^a = \sum_{i=1}^{N}\left(3(\mathbf{a}_i \cdot \mathbf{A})(\mathbf{a}_i \cdot \mathbf{B}) - \delta_{AB}\right)/2N$, where $\mathbf{a} = (\hat{x}, \hat{y}, \hat{z})$ is any of the molecular symmetry axes and $A, B = \hat{X}, \hat{Y}, \hat{Z}$ denote the axes of the simulation box. The onset of positionally ordered phases is usually accompanied by clear density jumps in the $p^* - \eta$ EOS. Visual inspection and combined analysis of the orientational/positional order through, the calculation of appropriate projections of the radial pair correlation functions,[21] allows the location of the phase transition pressures and an unambiguous classification of the positionally ordered states.

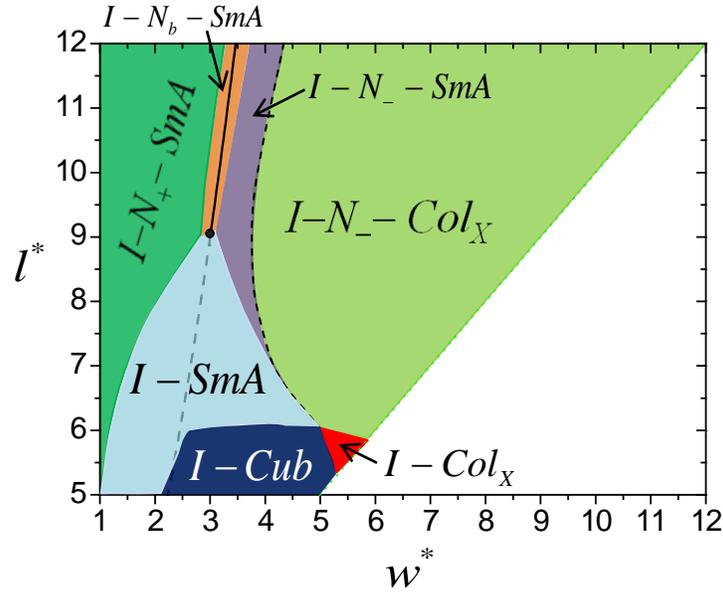

**Figure 2.** The phase diagram of hard spheroplatelets as a function of their molecular geometry parameters. Different colored regions correspond to different phase sequences. The boundaries between the regions are estimates inferred from the analysis of a large number of MC simulations at several SP molecular geometries ($l^*, w^*$). Straley's line is also plotted (black solid line). The black circle corresponds to the critical molecular geometry $(l_c^*, w_c^*) = (9, 3)$.

The mesophases of different ordering exhibited by the simulated systems in the range $5 \leq w^* < l^* < 12$ are: i) two uniaxial nematic phases, a calamitic ($N_+$) and a discotic ($N_-$), with the $z$- and the $x$- molecular axis aligned along the unique director respectively, ii) two biaxial nematic phases ($N_{b+}$ and $N_{b-}$) with the common alignment of the $z$- and $x$-molecular axis defining the primary director in the $N_{b+}$ and in the $N_{b-}$ respectively, iii) uniaxial $SmA$ phases with the long $z$-molecular axis oriented along the layer normal, iv) columnar phases ($Col_X$) either uniaxial or biaxial with the molecular $x$-axis oriented along the columns in all cases and v) a variety of cubatic ($Cub$) mesophases.

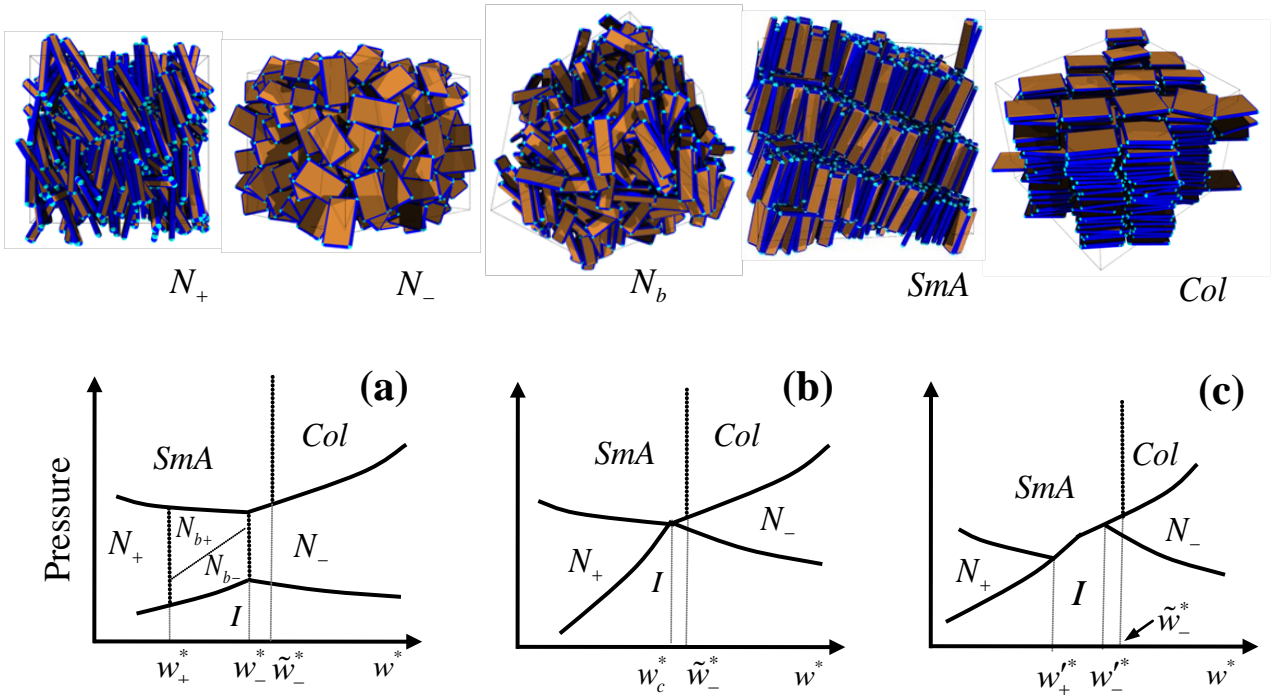

**Figure 3.** Pressure vs molecular width qualitative phase diagram topologies for molecular lengths: (a) $l^* > l_c^*$, (b) $l^* = l_c^*$, and (c) $l^* < l_c^*$. Representative snapshots of liquid crystalline phases are also shown.

Starting from the low pressure (packing fraction) isotropic state ($I$) and compressing the systems up to their crystallization or vitrification pressures we detected the following enantiotropic phase sequences: $I-[N_+]-SmA$, $I-[N_{b-}-N_{b+}]-SmA$, $I-N_--SmA$, $I-[N_-]-Col_X$ and $I-Cub$; the brackets here indicate phases that may be absent from the phase sequence. We present the complete phase diagram in the $w^* - l^*$ parameter space in Figure 2. Each differently colored domain corresponds to a different phase sequence. Clearly the phase diagram of Figure 2 is symmetric with respect to the $l^* = w^*$ line. The thick solid line (Straley's line) represents molecular geometries with $l^* = w^{*2}$ for which, according to theoretical considerations,[12,16] the particles are of "dual shape" --neither prolate nor oblate-- and biaxially ordered phases are expected.

The phase diagram of Figure 2 reveals that the $(l_c^*, w_c^*) \approx (9,3)$ point corresponds to a critical molecular anisometry. At this point the system exhibits a direct $I - SmA$ phase transition on compression. However, even a small deviation from this specific molecular anisotropy, leads to severe changes in the phase sequence/stability. To facilitate the discussion on the phase behavior of particles with dimensions close to the critical geometry we present in Figure 3 three distinct topologies of the $p^* - w^*$ phase diagram corresponding to SPs having reduced length $l^*$ slightly above (Figure 3a), equal (Figure 3b) and slightly below (Figure 3c) of the critical length $l_c^*$.

For lengths $l^* > l_c^*$ the first ordered phases observed upon compression are different types of nematics. The symmetry and structure of the nematic phase depends on the width $w^*$, see Figure 3a. Specifically, for $w^* < w_+^*$ a calamitic nematic ($N_+$) is obtained while for $w^* > w_-^*$ the phase is discotic ($N_-$). Here, $w_{(+/-)}^*$ are the critical SP widths that determine the limits of stability of the uniaxial nematic phases and vary with $l^*$. In both cases the $N - I$ phase transition has a weakly first order character. For SPs with widths in the narrow window $w_+^* < w^* < w_-^*$ a stable biaxial nematic phase enters into the phase sequence. The precise values of $w_{(+/-)}^*$ are difficult to be determined precisely from the simulations due to the almost second order nature of the $N - I$ phase transition in this window. To exclude the possibility that the biaxial nematic phases is an artifact due to the size of the simulated systems, much larger systems ($N \approx 4000$) where simulated that confirmed our findings. The long range nature of the biaxial order of the $N_b$ phases where confirmed with the calculation of a series of rotationally invariant pair correlation functions. Perhaps the most striking finding in this range of molecular geometries is the crossover $N_{b-} - N_{b+}$ between two distinct biaxial nematic states (in Figure 3a the crossover pressures are indicated with the thin dotted line that divides the biaxial window). These low and high pressure (packing fraction) biaxial nematics, although of the same macroscopic symmetry, differ in the molecular axis that determines the principal director of the phase. This packing fraction driven change of the main director of the nematic phase is expected to have profound implications on the alignment and/or the response of these phases upon the application of external aligning fields. Such a $N-N$ phase transition is supported by the Landau-deGennes theory for biaxial nematics[22], and was predicted for systems of biaxial particles with the help of Onsager type molecular theory by Taylor et. al.[23] and more recently confirmed by Martínez-Ratón et. al.[24] with a more elaborate density functional theory that goes beyond the second-order virial approximation.

All the nematic phases in the $l^* > l_c^*$ region transform upon compression through a first order phase transition, in either *SmA* or *Col* mesophases. The observation, however, that not only the $N_+$ phase but the biaxial and, for a narrow range of molecular widths $w_-^* < w^* < \tilde{w}_-^*$, the $N_-$ phase as well, transform into uniaxial (rod-like) *SmA* phases reveals a new class of phase transitions which, to the best of our knowledge, have not been reported previously for thermotropic or colloidal mesophases. Here, $\tilde{w}_-^*$ denotes the molecular width that determines the boundary between the $N_- - SmA$ and the $N_- - Col$ phase sequences.

The biaxial nematic states are not present in systems with molecular lengths $l^* < l_c^*$ (see Figures 2 and 3c). For these molecular elongations, the two uniaxial nematic phases are gradually reduced into the ranges $w^* < w_+'^*$ and $w^* > w_-'^*$ for the rod- and disc-like nematic phases respectively. In the intermediate range $w_+'^* < w^* < w_-'^*$ nematicity is totally suppressed and a direct first order phase transition from the isotropic to a uniaxial *SmA* phase is observed. For $l^* < 6$ the systems condense from the isotropic phase directly into i) the *SmA* phase for $w^* < 2.5$ or ii) cubatic mesophases for $w^* > 2.5$ or iii) the columnar phase for molecular widths $w^* \approx l^*$. Clearly, $l^* \approx 6$ corresponds to the lower molecular length bellow of which the nematic state disappears completely from the phase sequence.

For $6 < l^* < l_c^*$ the $N_-$ phase appears in the phase sequence for $w^* > w_-'^*$ and, as in the case with $l^* > l_c^*$, for a narrow range of molecular widths ($w_-'^* < w^* < \tilde{w}_-^*$) it transforms upon compression directly into the *SmA* phase . The $I - N_- - Col$ phase sequence is observed when $w^* > \tilde{w}_-^*$, (the $\tilde{w}_-^*$ boundary is indicated with the dashed line in Figure 2). In the region $w'^*_- < w^* < \tilde{w}_-^*$ the $N_-$ phase is thermodynamically stable for a small range of pressures (packing fractions). For these pressures the extent of *SmA* or *Col* ordering fluctuations in the simulated $N_-$ phases are comparable with the size of the simulation box and this makes difficult the precise determination of the $\tilde{w}_-^*$ -boundary. For molecular geometries close to the $\tilde{w}_-^*$ line the competition between *SmA* and *Col* ordering, in combination with the unavoidable molecular size polydispersity in real systems are expected to enhance the stability of the nematic state in favor of the positionaly ordered states.[13,25]

Our results suggest that at $l_c^* \approx 9$ (see Figure 3b) the lines $w_{+/-}^*$ and $w'^*_{+/-}$ converge onto the critical point $(l_c^*, w_c^*)$ at which the biaxial nematic phase, the two uniaxial nematics and the *SmA* phase meet. Interestingly the critical point $(l_c^*, w_c^*)$ lies on the Straley's line and defines, in terms of the molecular dimensions, the lower limit of stability of the biaxial nematic phase. In other words, according to our findings, the Straley's condition $l^* = w^{*2}$ for biaxial nematics is valid provided that $l^* > 9$.

In conclusion, performing a large number of MC computational experiments we were able to trace the phase boundaries of sterically interacting SPs, a prototype that mimics the very basic features of real biaxial board like colloids. Despite its simplicity, the model reproduces remarkably well many experimental observations on colloidal suspensions of board like particles. We have identified new types of phase transformations that include i) a $N_{b-} - N_{b+}$ crossover and ii) the direct transition from a discotic nematic or a biaxial nematic to a conventional orthogonal *SmA* mesophases. Our results provide a robust framework and a well defined reference system for understanding and interpreting the presence of more complex interactions in real colloids upon detecting deviations from the hard core model. These nearly exact computational experiments on the phase behavior of hard biaxial particles offer a comprehensive guide to experiments, towards the design of colloidal systems with the desired functionality, as well as to theory for testing and improving analytical molecular models using simple intermolecular potentials.

ACKNOWLEDGMENT

This research has been co-financed by the European Union (European Social Fund –ESF) and Greek national funds through the Operational Program "Education and Lifelong Learning" of the National Strategic Reference Framework (NSRF)-Research Funding Program: THALES.